\documentclass[12pt,preprint]{aastex}
\usepackage{natbib,amssymb}

\bibpunct{(}{)}{;}{a}{}{,} 
\usepackage{graphics,graphicx} 

\newcommand{\emp}{Earth, Moon Planets} 
\newcommand{\jncs}{J.~Non-Cryst.~Solids} 
\newcommand{\mps}{Meteoritics and Planetary Science} 
\newcommand{\sci}{Science}             

\shorttitle{Crystallinity of interstellar silicates}
\shortauthors{F.~Kemper et al.}
\begin{document}
\newcommand{\ud}{\mathrm{d}}

\title{The absence of crystalline silicates in the diffuse interstellar medium\footnote{Based on observations with ISO, an
    ESA project with instruments funded by ESA Member States
    (especially the PI countries: France, Germany, the Netherlands and
    the United Kingdom) and with the participation of ISAS and NASA}}

\author{F.~Kemper\altaffilmark{1,2}, W.J.~Vriend\altaffilmark{3} and A.G.G.M.~Tielens\altaffilmark{3,4}}
\email{kemper@astro.ucla.edu}

\altaffiltext{1}{Department of Physics and Astronomy, University of California Los Angeles, 405 Hilgard Avenue, Los Angeles, CA 90095-1562}

\altaffiltext{2}{Spitzer Fellow}

\altaffiltext{3}{Kapteijn Institute, University of
Groningen, P.O.~Box 800, 9700 AV Groningen, The Netherlands}

\altaffiltext{4}{SRON
Laboratory for Space Research, P.O.~Box 800, 9700 AV Groningen, The
Netherlands}

\begin{abstract}
Infrared spectroscopy provides a direct handle on the composition and
structure of interstellar dust.  We have studied the dust along the
line-of-sight towards the Galactic Center using Short Wavelength
Spectrometer (SWS) data obtained with the Infrared Space Observatory
(ISO). We focussed on the wavelength region from 8--13 $\mu$m which is
dominated by the strong silicate absorption feature. Using the
absorption profiles observed towards Galactic Center Sources (GCS) 3
and 4, which are C-rich Wolf-Rayet Stars, { as reference objects,}
we are able to disentangle the interstellar silicate absorption and
the silicate emission intrinsic to the source, toward Sgr A$^*$ and
derive a very accurate profile for the intrinsic 9.7 $\mu$m band. The
interstellar absorption band is smooth and featureless and is well
reproduced using a mixture of 15.1\% amorphous pyroxene and 84.9\% of
amorphous olivine { by mass}, all in spherical sub-micron-sized
grains. There is no direct evidence for substructure due to
interstellar crystalline silicates. By { minimizing} $\chi^2$ { of spectral fits
to the absorption feature}
we are able to determine an upper limit to { the degree of
crystallinity of silicates} in the diffuse interstellar medium (ISM),
and conclude that the { crystalline fraction of the interstellar
silicates} is 0.2\% $\pm$ 0.2\% by mass.  { This} is much lower
than the degree of crystallinity observed in silicates { in the circumstellar
environment of}
evolved stars, { the main contributors of dust to the ISM}. {
There are two possible explanations for this discrepancy.} First, an
amorphization process  occurs in the { ISM} on a time scale
significantly shorter than the destruction time scale, possibly caused
by { particle bombardment by} heavyweight ions. Second, we consider the
possibility that the crystalline silicates in stellar ejecta are
diluted by an additional source of amorphous silicates, in particular
supernovae.  { We also compare our results with a study on silicate
pre-solar grains found in interplanetary dust particles.}
\end{abstract}


\section{Introduction}

In the last decade, a multitude of evidence for the presence of
crystalline silicates in various astrophysical environments has
emerged. In particular, infrared spectra have revealed that silicates
in circumstellar environments often contain a significant crystalline
fraction, both around post-main-sequence stars
\citep[e.g.][]{WMJ_96_mineralogy,MWT_02_xsilI} and pre-main-sequence
stars \citep[e.g.][]{WWD_96_xsilsyoung,MWB_01_haebe}. In addition,
crystalline silicates are ubiquitous in the Solar System, not only in
the more evolved bodies such as planets, but also in primitive objects
like comets \citep[e.g.][]{W_02_cometgrains}. Because crystallization
is inhibited by high energy barriers, the origin and evolution of the
crystalline silicate fraction in interstellar and circumstellar media
has the potential to provide direct evidence of the energetic
processing of grains.  

The life cycle of dust starts in the outflow of evolved stars,
continues upon ejection in the interstellar medium and eventually ends
in the planet forming disk around a young star.  It is surprising
that, whereas at the beginning and end of dust grains' lifes
crystallinity is prevalent, no crystallinity is found in the
intermediate phase (i.e.~in the diffuse interstellar medium). In fact,
a relatively high upper limit to the degree of crystallinity in the
diffuse ISM has been determined recently
\citep{LD_01_silicate}. Less than 5\% by number of the interstellar
Si-atoms were found to be incorporated in crystalline silicate grains
of $<1$ $\mu$m in size, which is roughly equivalent to a mass fraction
of $< 5$\%. On the other hand, \citet{BA_02_mineralogy} have suggested
that, while the broad and structureless interstellar 10 $\mu$m
absorption feature is commonly ascribed to amorphous silicates, the
crystalline spectral detail may be washed out in a very complex
mixture of crystalline silicates. Furthermore, some studies on
silicates in the dense ISM have reported the (controversial) detection
of crystalline silicate features
\citep{CJL_00_xsil_in_orion,OO_03_carbononions}, but one has to bear in mind that
this environment has very different physical properties than the
diffuse ISM.

In this work, we will re-address the issue of crystallinity in the
diffuse ISM by studying the line-of-sight towards the Galactic
Center. Because of its large amount of extinction and its high
infrared flux, the sightline towards the Galactic Center has often
been used to characterize the properties of interstellar dust
\citep{RA_85_extinction,RRP_89_gc,PSA_94_GC,TWA_96_GC,LFG_96_gc,CPG_98_hydrocarbon}.
We will study the 10 $\mu$m silicate absorption feature in order to
determine the degree of crystallinity in this line-of-sight. In
Sect.~\ref{sec:obs} we will discuss the ISO SWS observations and data
reduction, as well as the characteristics of the region around Sgr
A$^{*}$ and the correction method applied for emission 
intrinsic to the Galactic Center (GC) region. The method used to
determine the dust composition in the diffuse ISM is described in
Sect.~\ref{sec:determination} along with the results. In
Sect.~\ref{sec:processing} we discuss two mechanisms to explain the
discrepancy in crystallinity observed between stellar ejecta and the
diffuse ISM. A comparison with silicates in the solar system and
planet forming disks around other stars is given in
Sect.~\ref{sec:comparison}. Sect.~\ref{sec:conclusions} contains the
conclusions.

\section{Observations and data reduction}
\label{sec:obs}

\subsection{ISO SWS observations of Sgr A$^{*}$ and nearby objects}

During the life time of the ISO mission \citep{KSA_96_ISO}, 
2.38 -- 45.2 $\mu$m SWS spectra \citep{GHB_96_SWS} of the Galactic
Center and { two sight lines in the nearby Quintuplet cluster} were obtained. Table~\ref{tab:obs} gives an
overview of these observations.

\begin{figure}
\plotone{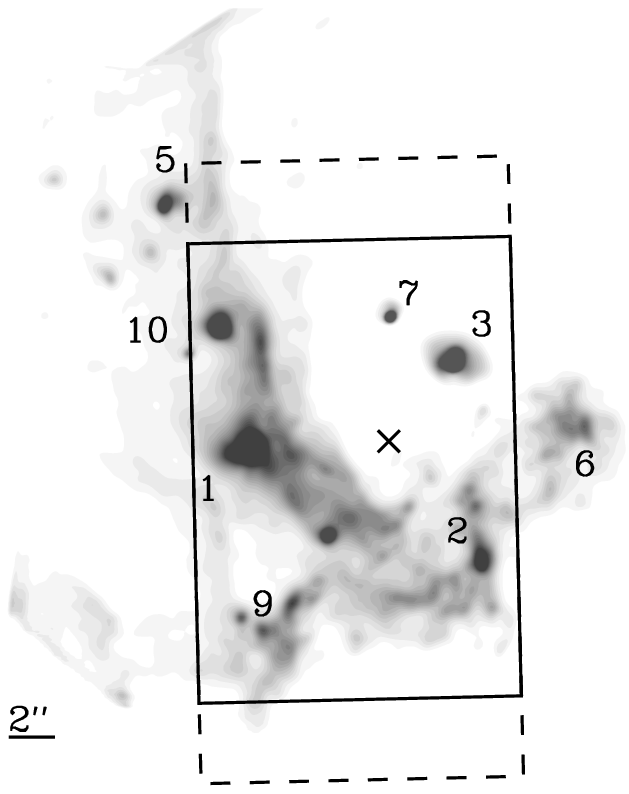}
\caption{The position of the ISO SWS beam toward the Galactic Center.
This plot shows a map of the Galactic Center at { 12.4 $\mu$m
\citep{TGM_02_GC}, on which we overplotted the pointing of ISO SWS. The position of Sgr A$^{*}$ is
indicated with a $\times$, and the numbers indicate the positions of
the GC IRS sources, following the notation from
\citet{BMN_78_GCI}. The beam for the SWS detectors working at $\lambda
\le 12.0$ $\mu$m is indicated with a solid line. The dashed line
indicates the beam size when observing at $12.0 < \lambda \le 27.5$
$\mu$m. The beam includes infrared sources 1, 2, 3, 7, 9, 10.  For an
image showing the two ISO SWS beam positions toward the Quintuplet
sources, the reader is referred to Fig.~1 of
\citet{MSB_01_quintuplet}.}}
\label{fig:beam}
\end{figure}

The Galactic Center { region} is very crowded, while the beam
ISO used for the SWS observations is very large. In
Fig.~\ref{fig:beam} the orientation and positions { of the beam is}
indicated { on a 12.4 $\mu$m map of the Galactic Center
\citep{TGM_02_GC}}. { The beam}
includes infrared sources (IRS) 1, 2, 3, 7, 9 and 10, of which IRS 1
is the brightest at 12.4 $\mu$m. At the position of Sgr A$^*$ itself
there is virtually no 12.4 $\mu$m emission. IRS sources 1, 2, 9 and 10
lie along an arc on the so-called Ridge, generally believed to be a
complex of H{\sc ii} regions. IRS 7 is a late-type supergiant and is
the brightest 2.2 $\mu$m source in the vicinity of the galactic
center. IRS 3, an OH/IR star, shows the 10 $\mu$m silicate feature in
absorption \citep{BMN_78_GCI}. In the large SWS beam, the mid-IR
emission is dominated by the extended emission associated with IRS 1.
 
GCS 3 and 4 are members of the Quintuplet cluster.  The positions of
the beams are indicated on the ISOCAM image presented { in Fig.~1
of the study by \citet{MSB_01_quintuplet}}. In the observation of GCS
3, the beam was centered on GCS 3-I { (or Q4)}, however, GCS 3-II { (or Q2)} is the brightest {
mid-infrared} source in the beam and dominates the measured spectrum
\citep{MSB_01_quintuplet}.

\subsection{Data reduction}

The full ISO SWS spectrum of the line-of-sight towards Galactic Center
was first published by \citet{LFG_96_gc}. We chose to reduce this
spectrum again to fully address the problems caused by the detector
memory effects in band 2. In addition, we used the SWS spectra of GCS
3 and 4 { as a reference} to correct for the silicate emission
intrinsic to { the Sgr A$^{*}$ region}. The spectra of GCS 3 and 4 were previously
published in { the fore-mentioned work} by
\citet{MSB_01_quintuplet}, but we redid the data reduction to be
consistent with the reduction of the spectrum of Sgr A$^*$.

We used the Interactive Analysis package (IA$^3$) to reduce all three
spectra. In addition to the automated routines provided in this
package, the up and down scans were analyzed separately to remove
glitches and detector jumps by hand. To correct for the detector
memory effects, most prominently present in band 2 (ranging from 4.08
to 12.0 $\mu$m), the Fouks-Schubert model was applied (see
Sect.~\ref{sec:memory}). From the spectra in band 2c and all sub-bands
of band 3 (ranging from 7.00 to 29.0 $\mu$m) we removed residual
fringes which were present after the Relative Spectral Responsivity
Function (RSRF) was applied. Rebinning according to actual spectral
resolution was performed ($R=1000$ for the speed 4 observations of Sgr
A$^*$; $R=500$ for the speed 3 spectra of GCS 3 and 4), and data
points deviating more than 3$\sigma$ per bin were removed, resulting
in a loss of 2\% of the data points. Narrow spectral lines were
excluded from sigma-clipping. Finally the spectra from the individual
detectors were rescaled to the average flux levels to produce a smooth
spectrum rather than a dot cloud.

\subsection{Detector memory effects}
\label{sec:memory}

It is known that ISO SWS spectroscopy suffers from memory effects in
the data, a problem predominantly found in data from bands 2 and 4
(i.e.~the wavelength ranges from 4.08--12.0 $\mu$m and 29.0--45.2
$\mu$m). The semi-conductors that the detectors consist of do not
fully discharge between subsequent measurements in a wavelength
scan. When the flux levels in a certain (sub-)band vary only by a
small amount, this is not a large problem, since the amount of memory
in each measurement will be more or less constant, and simply
generates an offset in the flux levels. The relative flux levels will
not be affected, so it may be relatively easy to correct for the
memory when subtracting the
\emph{dark current}. In cases where a large variation in flux levels
occurs within one sub-band, the memory effects require additional
attention. This is the case in band 2c, ranging from 7.00--12.0
$\mu$m, which contains the entire interstellar absorption feature due
to silicates. The discharging of the semi-conducting material can be
described with a set of differential equations
\citep{FS_95_memory}. This model expresses the time delay in the
charging on the incident intensity. The larger the change in flux
levels, the longer it takes to reach the charging corresponding to the
incident flux. \citet{V_99_10micronGC} has applied the Fouks-Schubert
model to the data in band 2, and was able to correct for the memory
effects. The successful application of this model can be easily
checked by comparing the data of the up- and down-scans separately,
since the memory effects should affect these data in opposite
wavelength directions. This difference is minimized when the memory
correction is applied successfully. Only a very small residual memory
effect remains.  The application of the Fouks-Schubert model to
correct for memory effects in band two, which was first developed for
the data analysis of the GC line-of-sight
\citep{V_99_10micronGC}, is now implemented in the IA$^3$
package and the ISO standard pipe line data products
\citep{LKS_02_ISOSWS,K_03_memory}.

\subsection{Correction for silicate emission intrinsic to Sgr A$^{*}$}

\begin{figure}
\epsscale{0.6}
\plotone{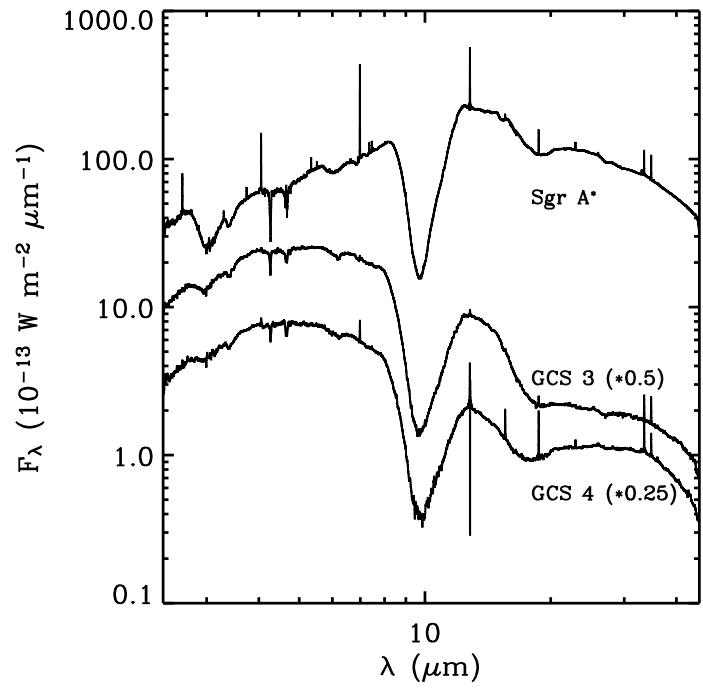}
\caption{The 2.38--45.2 $\mu$m ISO SWS spectra of Sgr A$^{*}$, GCS 3 and GCS 4. To
enhance the clarity of the figure, the flux levels of GCS 3 and GCS 4
are multiplied with a factor of 0.5 and 0.25 respectively.}
\label{fig:spectra}
\end{figure}

\begin{figure}
\epsscale{0.6}
\plotone{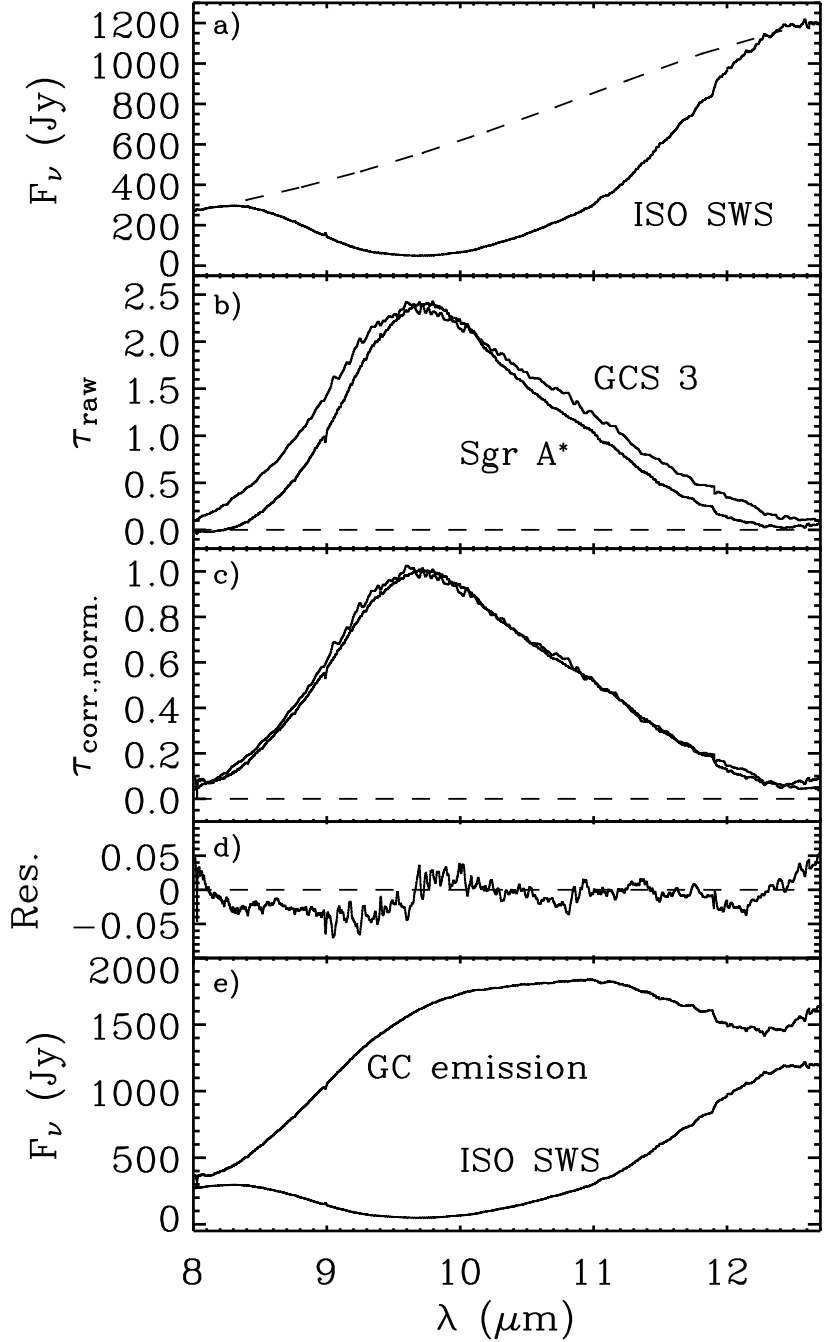}
\caption{Steps in the data reduction process. { Panel a) shows the observed spectrum
of the Sgr A$^*$ region, along with a fourth order polynomial
continuum (dashed line). Panel b) shows the raw optical depth in the
10 $\mu$m feature in Sgr A$^*$ and GCS 3. In panel c) the Sgr A$^*$
absorption feature is  corrected for emission intrinsic to the
Galactic Center, and now nicely overlaps with the 
optical depth towards GCS 3. Both spectra are normalized. The
remaining difference between the two normalized optical depths is
shown in panel d). Finally, panel e) shows what the intrinsic emission
from the GC region would have looked like if there were no
interstellar silicate absorption. This part of the spectrum is only
de-reddened for optical depth \emph{in} the silicate feature, however
this feature is of course superposed on a continuum of extinction,
giving rise to $\sim$30 magnitudes of extinction in the optical.}}
\label{fig:mem}
\end{figure}

Fig.~\ref{fig:spectra} shows the ISO SWS spectra of Sgr A$^{*}$, GCS 3
and GCS 4. Since it is thought that the absorption feature at 10
$\mu$m is caused by interstellar dust particles along the
line-of-sight, and that the dust composition along those
lines-of-sight is very similar, it is remarkable that the 10 $\mu$m
absorption feature towards Sgr A$^{*}$ seems to be much narrower than
the features observed towards the Quintuplet stars. { This is
illustrated in panel b) of Fig.~\ref{fig:mem}, which gives the raw
optical depth $\tau_{\mathrm{raw}}$ for GCS 3 and Sgr A$^*$ calculated
directly from $F_{\nu,\mathrm{obs}} = F_{\nu,0}
e^{-\tau_{\mathrm{raw}}}$, where $F_{\nu,0}$ is a continuum determined
using fourth order polynomial fitting.}

The stars in the Quintuplet cluster are identified as carbon-rich
Wolf-Rayet stars \citep{FMM_99_quintuplet}. We therefore assume that
there are no circumstellar silicates that give rise to emission or
absorption features intrinsic to the Quintuplet stars and that the
shape of the feature is entirely determined by interstellar
absorption. Indeed,
\citet{MSB_01_quintuplet} conclude that the Quintuplet cluster does not contain 
any gas or dust between its stellar components and that the silicate
absorption feature is a truly interstellar feature.

We assume that the dust composition in the line-of-sight towards Sgr
A$^*$ does not differ much from that towards the Quintuplet sources,
and therefore conclude that the difference of the absorption profile
can be explained by silicate emission intrinsic to the Sgr A$^*$
region. It has been reported that the line-of-sight towards the GC
contains more molecular clouds than those toward the Quintuplet
sources \citep{CTW_01_GC}. We assume that this only leads to a
difference in optical depth, and not to a difference in dust
composition.  Rather, we attribute the difference in the 10 $\mu$m
profile of Sgr A$^{*}$ and the Quintuplet sources to the presence of
underlying silicate emission in the former. Since the S/N of the
spectrum observed towards Sgr A$^*$ is much better than the data of
the Quintuplet source, we are aiming to correct for the silicate
emission local to Sgr A$^*$ to obtain the intrinsic interstellar
profile. We assume that the silicate contribution local to Sgr A$^*$
gives rise to the same feature in emission as it does in
absorption. Likely, the silicate emission is associated with dust in
the H{\sc ii} region, IRS 1. { Because the resulting observed
absorption feature is determined by the intrinsic emission and
interstellar absorption, we applied an iterative process to determine
the shape of the interstellar absorption and GC emission feature.
The differences with GCS 3 or GCS 4
are minimized (the best results are achieved for GCS 3, see panel c)
and d) of Fig.~\ref{fig:mem}), finally resulting in the corrected
interstellar absorption feature. Panel e) of Fig.~\ref{fig:mem} shows
the spectral appearance of the GC silicate emission feature, where the
reader must keep in mind that: First, the emission is the sum of all
the mid-infrared sources in the ISO beam, and second, the spectrum is
only de-reddened for the optical depth \emph{in} the feature and that the fact
that the silicate interstellar absorption feature is superposed on a
extinction continuum is ignored.}

A similar correction { to determine the interstellar absorption in
the 10 $\mu$m feature} was first performed by
\citet{RA_85_extinction}, who used the silicate profile of red supergiant
$\mu$ Cep as a template for the interstellar absorption in order to
de-redden their observations of the galactic center environment. This
was justified by the resemblance between the $\mu$ Cep silicate
feature and the absorption feature observed towards a number of
carbon-rich Wolf-Rayet stars \citep{RA_84_WCWR}.

\section{Determination of the dust composition}
\label{sec:determination}

\subsection{Method}
\label{sec:method}

\begin{figure}
\epsscale{0.5}
\plotone{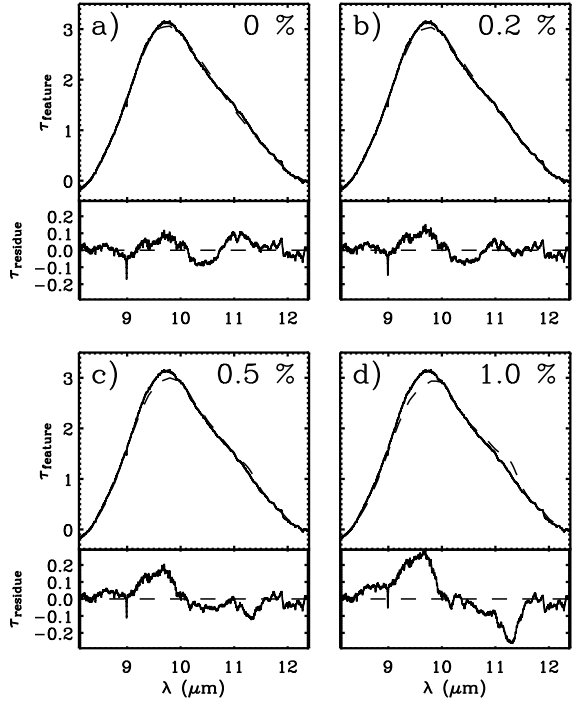}
\caption{Optical depth observed in the 10 $\mu$m silicate feature towards
Sgr A$^*$. Panel { a)} shows the best $\chi^2$ fit consisting of a
mixture of amorphous silicates (dashed line) to the optical depth in
the feature (solid line). The lower part of panel { a} shows the residual optical
depth after the fit is subtracted from the observed optical depth. 
Panels { b)}, { c)}, { d)} show the best $\chi^2$ fit (dashed line) to
the ISO data (solid line) of partially 
crystalline mixtures of silicates with the same ratio of pyroxenes over olivines
as in panel { a)}. In each panel the degree of crystallinity is indicated
in the upper right corner, and the lower part of each panel shows the residual
optical depth. In case of the completely amorphous dust composition, the
residuals are at most $\sim$ 3\% of the optical depth of the amorphous
silicates. The $\chi^2$ values are smallest for the 0.2\% degree of 
crystallinity, which is evident from the residuals as well. With increasing 
crystallinity, the fit quickly 
deteriorates, which becomes visible as larger residues. 
At $\sim$0.5\% the fit is { already worse than} a completely amorphous fraction.  }
\label{fig:tau}
\end{figure}

{ The corrected optical depth $\tau$ is presented in
Fig.~\ref{fig:mem}c and Fig.~\ref{fig:tau} and can be compared
directly to} the opacity of the various dust components at these
wavelengths.  The mass absorption
coefficient $\kappa$ correlates to the optical depth according to
$\tau_{\lambda} = \rho_{\mathrm{d}} \kappa_{\lambda} L = n
m_{\mathrm{d}} \kappa_{\lambda} L$, where $L$ is the distance to
the galactic center, $\rho_{\mathrm{d}}$ is the average density of the
considered dust species, { $n$ is the number density of grains and
$m_{\mathrm{d}}$ the average mass of a dust grain. We can define
a the mass column density $N_i = \rho_{\mathrm{d},i} L$ and thus fit the 10$\mu$m
absorption feature by solving:}

\begin{equation}
\tau_{\lambda} = \sum_{i} N_i \kappa_{\lambda,i}
\label{eq:tau}
\end{equation} 

\noindent { for $N_i$}. A continuum subtraction is performed on the mass absorption
coefficients between the same wavelength boundaries which were used to
obtain the optical depth in the interstellar 10 $\mu$m
feature. Consequently only the contribution to the optical depth in
features is taken into account.

In order to solve Eq.~(\ref{eq:tau}) we applied the $\chi^2$ fitting
method, which involves minimizing

\begin{equation}
\chi^2 = \sum_{\lambda} \Big( \frac{\tau_{\lambda} - \sum_{i} N_i \kappa_{\lambda,i}}{\sigma_{\lambda}} \Big)^2
\label{eq:chisq}
\end{equation}

\noindent where the $\sigma_{\lambda}$ is the uncertainty in the measured values
arising from our data reduction procedure. They represent the spread
in individual measurements of the twelve different SWS detectors,
before the spectrum was collapsed into one point per spectral bin. A
measure for the goodness of the fit is given by the reduced $\chi^2$
value, $\chi_\nu^2$, { defined as} $\chi_\nu^2 = \chi^2 / \nu$ with
$\nu = N_p - m$. The parameter $N_p$ represents the number of data
points in the spectrum, and $m$ gives the number of free parameters.
The $\chi_\nu^2$ values derived by our fitting method can be very high
(at least $\sim 44$), while textbook examples always require this
value to be close to unity. Two different effects contribute to these
high values.

First, $\sigma_{\lambda}$ is only determined by the random scatter in
the measurements, while the systematic errors in the flux levels can
be significantly higher than that. Sources for these systematic errors
could be residuals of the memory effect, which can be as high as 2\%
in either direction or features introduced by the responsivity
functions. Second, the application of a $\chi^2$ fitting method
assumes that the model does not contain uncertainties. In fact the
model spectra we use are based on mass absorption coefficients
measured in the laboratory and therefore have intrinsic
uncertainties. It is hard to quantify the combined effect of the
uncertainties in the laboratory spectra and the systematic errors in
the ISO SWS data. We have elected to use only the uncertainties
directly resulting from the ISO SWS data in the $\chi^2$ fitting
method.

We performed fits with different dust components, both amorphous and
crystalline silicates.  First, we have determined confidence levels
for the fit with only amorphous dust components (see
Sect.~\ref{sec:amorphous}) using the method discussed by
\citet{PTV_92_recipes}.  In addition we use the $F$-test
\citep{BR_92_ftest} to determine the goodness of the fits when
crystalline silicates were added. This test shows how much a $\chi^2$
fit improves if a parameter is added. The value $F_\chi$ is calculated
from

\begin{equation}
F_\chi = \frac{\Delta \chi^2}{\chi_\nu^2}
\end{equation}

\noindent where $\Delta \chi^2$ is the decrease in $\chi^2$ between the initial
and new fit, and $\chi_\nu^2$ is the value from the fit with the extra
parameter added (i.e.~the new fit). The parameter $F$ measures the
relative improvement of a the $\chi^2$ of a fit to randomly scattered
data when an additional component is included. The larger $F$, the
more significant is the improvement \citep[see e.g.~p.~209
of][]{BR_92_ftest}.

\subsection{The amorphous component}
\label{sec:amorphous}

In order to { find} the mass fraction contained in the crystalline
silicates, the total mass in the amorphous component { has to} be
determined, { which depends on its composition}. The chemical
composition of amorphous silicates { is determined by} the
formation path. Amorphous silicates resulting from the amorphization
of crystalline silicates will have the stoichiometry of that crystal,
while amorphous silicates formed through direct condensation from a
gas cloud or by { collisions} and subsequent merging of two { distinct}
grains are most likely to have a non-stoichiometric composition. Most
measurements of the optical properties of amorphous silicates are
limited to the stoichiometries of olivines and pyroxenes, except for a
recent study by \citet{JDM_03_solgel}, who applied the sol-gel method
to construct amorphous silicates of some selected non-stoichiometric
compositions. { In the sol-gel method Mg- and Si-hydroxides
(Mg(OH)$_2$ and Si(OH)$_4$) are mixed in the liquid phase. Extraction
of H$_2$O by chemical reaction between the Mg- and Si-hydroxides then
gives rise to the formation of silicates. Because the two components
can be mixed in any desired ratio, it is possible to synthesize
non-stochiometric amorphous Mg-silicates. For more details, the reader
is referred to the work by
\citet{JDM_03_solgel}. However, the
spectral appearance of these materials does not match the interstellar absorption
feature, therefore we will only consider stoichiometric compositions.} We
have used the optical constants determined by
\citet{DBH_95_glasses} for olivine (MgFeSiO$_4$) and { pyroxene}
(MgFeSi$_2$O$_6$). { The use of laboratory optical constants allows us to immediately
convert the derived optical depths into column densities of dust as well 
as derive their chemical composition.}

\begin{figure}
\plotone{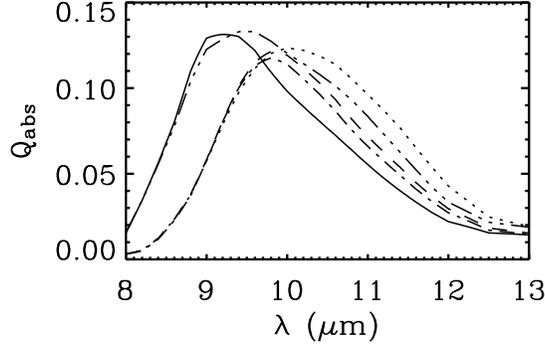}
\caption{Absorption efficiencies of amorphous silicates in the 8--13 $\mu$m range, 
calculated from optical constants provided by
\citet{DBH_95_glasses}. The absorption efficiencies of amorphous olivine (MgFeSiO$_4$)
in the form of spherical 0.5 $\mu$m-sized grains are indicated with a
dashed line, while the spherical grains with sizes $\lesssim$ 0.1
$\mu$m are indicated with the dashed-dotted line. The dotted line
represents non-spherical amorphous olivine grains. The solid line
gives the absorption efficiencies of spherical 0.1 $\mu$m-sized
pyroxene (MgFeSi$_2$O$_6$), and the dashed-triple-dotted line indicates
the values for non-spherical pyroxene grains.}
\label{fig:asils}
\end{figure}

The appearance of the 10 $\mu$m feature depends on the grain shape and
grain size of the silicate particles. In Fig.~\ref{fig:asils} an
overview of the most important effects is given, for both amorphous
olivine and pyroxene. Small spherical grains (up to $\sim 0.1$ $\mu$m
in size) all produce exactly the same shape for the spectral
feature. For larger grains, the feature starts to broaden towards
longer wavelengths. This effect becomes stronger with increasing grain
size. The use of non-spherical grains, represented by a continuous
distribution of ellipsoids
\citep[CDE;][]{BH_83_scattering}, tends to shift the peak of the
feature to longer wavelengths. With CDE calculations it is impossible
to include any information on grains size.  Amorphous olivine in the
form of spherical grains small in size compared to the wavelength (the
Rayleigh limit) provide a good fit to the 10 $\mu$m feature
\citep{V_99_10micronGC}. It is even possible to exclude the presence of a
high fraction of larger grains as well as more than 1\% of the mass
in the form of forsterite, based on this spectrum
\citep{BMD_01_processing}. We find however, that the fit using just amorphous
olivine can be further improved by adding amorphous pyroxene (also in
small spherical grains). This is consistent with the suggestion that
part of the silicates in the ISM are converted from olivines into
pyroxenes to explain the pyroxene presence in protostars
\citep{DDW_00_nearinfraredproblem}.  

Our result that spherical grains smaller than $\sim 0.1$ $\mu$m are
responsible for the 10 $\mu$m absorption feature is more or less
consistent with the ISM grain size distribution determined by
\citet[MRN]{MRN_77_grainsize}, which has an upper limit of 0.25
$\mu$m. Based on a ROSAT study of the X-ray scattering halo of Nova
Cygni 1992,
\citet{WSD_01_Xray} found that they needed to extend the MRN grain size
distribution to sizes much larger than 0.25 $\mu$m in order to explain
the observed scattering halo. In contrast, the same observations led
\citet{DT_03_Xray} to conclude they did not need such large
grains; grains with sizes larger than 0.4 $\mu$m accounted for only a
small amount of the observed scattering. Still a significant
contribution to the scattering is caused by the 0.1--0.4 $\mu$m sized
grains, while these grains do not really contribute to the 10 $\mu$m
silicate absorption profile discussed in this study.  { Of course,} the
determination of grain sizes through X-ray scattering is weighted
towards larger grain sizes, since those grains dominate the forward
scattering cross section, while the infrared absorption profile is
determined by the volume weighted average of the silicate grain size
distribution, { putting less weigth on this discrepancy between the IR and X-ray results.}

The best fit (lowest $\chi^2$ value) of the 10 $\mu$m feature using
small spherical olivine and pyroxene grains is given in panel { a)}
of Fig.~\ref{fig:tau}. We find that olivine contributes 84.9 \% by
mass and pyroxene 15.1 \%, for a fit of the 10 $\mu$m feature between
8.3 and 12.3 $\mu$m.  The remaining residues are at most of the order
of $\tau \approx 0.1$ which is about 3 \% of the optical depth in the
10 $\mu$m feature itself. { These results were achieved using only
the amorphous silicates with number ratio Mg/Fe =
1. \citet{DBH_95_glasses} have measured the optical properties of
amorphous pyroxenes Mg$_x$Fe$_{(1-x)}$SiO$_3$, with x=0.4, 0.5, 0.6,
0.7, 0.8, 0.95, and 1.0 and olivines Mg$_{2x}$Fe$_{2(1-x)}$SiO$_4$
with x=0.4 and 0.5. The synthesis of silicates with values of $x$
outside the listed range (i.e. $x \le 0.3$ for the pyroxenes and $x \le
0.3$ or $x \ge 0.6$ for the olivines) resulted in the formation of
crystalline rather than amorphous silicates (C.~J\"ager,
\emph{priv.comm.}). The spectral differences within the pyroxene and
olivine family are smaller than the differences between those
families. We have determined the $\chi^2$ and $\chi^2_\nu$ values for
several amorphous silicate mixtures, and found that only a few
combinations led to an improvement according to the $F$-test. The
$F$-values are only modest (of the order of 200 at most) compared to
the improvements achieved when the crystalline silicates are added
(see Tab.~\ref{tab:f}), and indicate that the pyroxenes are probably
slightly Mg-rich, with a composition between
Mg$_{0.5}$Fe$_{0.5}$SiO$_3$ and Mg$_{0.6}$Fe$_{0.4}$SiO$_3$. Although
the sampling of the olivines is limited, the fit to the 10 $\mu$m
feature improves marginally ($F_\nu
\approx$ 100) when Mg$_{0.8}$Fe$_{1.2}$SiO$_4$ is added to the
mixture of dust components, indicating a slightly Fe-rich
composition. Since the improvements are very small, and the
composition is very close to a Mg/Fe ratio of unity, we will consider
only the amorphous silicates with $x=0.5$ in the remainder of this
analysis. }

In case of the amorphous silicates, we have solved Eq.~(\ref{eq:tau})
using extinction efficiencies $Q$ rather than mass absorption
coefficients $\kappa$. We find that for the given dust composition and
wavelength interval over which the fit is performed the derived mass
fractions of 84.9 \% and 15.1 \% for olivine and pyroxene respectively
are accurate within 0.1\% with a confidence of more than 99.99\%. In
other words, the ratio olivine:pyroxene = 5.6:1.  However, when
crystalline silicates are added (see Sect.~\ref{sec:crystalline}), we
find that the ratio between amorphous olivines and pyroxenes may vary,
from 5.6:1 to 4.8:1.

Based on this fit, we can not only determine the relative masses of
olivine and pyroxene, but also the column densities of the amorphous
silicates towards the Galactic Center, and hence the silicate density
in the diffuse ISM averaged over this line of sight.

\subsubsection{The abundance of Si}

The silicate { mass column} density $N_{\mathrm{sil}}$ in the diffuse ISM is given
by

\begin{equation}
N_{\mathrm{sil}} = \left( \frac{4\, a\, \rho_{\mathrm{s}}}{3} \right)
\left(\frac{\tau_{\mathrm{9.7}}}{Q_{\mathrm{9.7}}} \right)  
\end{equation}
 
with $\rho_{\mathrm{s}}$ being the specific density of the mineral the
grains are made of. The continuum subtracted optical depth is given by
$\tau_{\mathrm{9.7}}$, while $Q_{\mathrm{9.7}}$ is the continuum
subtracted extinction efficiency. The parameter $a$ gives the radius
of the grain, and the distance to Sgr A$^{*}$ is represented by $L$.
The densities $\rho_{\mathrm{s}}$ are 3.71 g cm$^{-3}$ and 3.2 g
cm$^{-3}$ for olivine and pyroxene respectively
\citep{DBH_95_glasses}.  
{ We find that the mass column density of olivines in the diffuse
ISM $N_{\mathrm{oliv}} = 1.3 \times 10^{-3}$ g cm$^{-2}$. For the
pyroxenes we find $N_{\mathrm{pyr}} = 2.2 \times 10^{-4}$ g
cm$^{-2}$. These two interstellar silicate components add up to a
total column density of $N_{\mathrm{sil}} = 1.5 \times 10^{-3}$ g
cm$^{-2}$.}

{ Under the assumption of solid grains we find that the column
density of Si-atoms contained in silicates in the diffuse
line-of-sight toward the Galactic Center is
$N(\mathrm{Si}\,\mathrm{in}\,\mathrm{silicates}) = 5.6 \times 10^{18}$
cm$^{-2}$. A recent estimates for the column densities of neutral
hydrogen based on X-ray scattering yields $N(\mathrm{H}) = 1 \times
10^{23}$ cm$^{-2}$
\citep{BMM_03_xrayGC}. Adopting this value, the abundance of Si in silicates is
$\bigg[ \frac{N(\mathrm{Si}\,\mathrm{in}\,\mathrm{silicates})}{N(\mathrm{H})} \bigg]
= -4.25$, which is somewhat higher than the \emph{total} Si-abundance
of $\sim -4.4$ found in the solar neighborhood
\citep{SCS_94_abundance}.  Porosity of the silicate grains
is suggested as an explanation, because porous grains appear to have
stronger 9.7 $\mu$m resonances relative to $A_V$ \citep{M_98_porous,ICW_01_porous}.
This is supported by observations of the ratio
$A_V/\tau_{9.7} \approx 19$  averaged over several
sight lines \citep[e.g.][]{RA_84_WCWR,M_98_porous}, while this ratio
seems to be $\sim 10$ towards the Galactic Center. 
On the other hand, it seems unlikely
that grain properties in GC line-of-sight are different from
other sight lines, and a more likely explanation may be found in the
enhanced metallicity in the inner region of the Galaxy, which could
easily explain the slight overabundance observed towards Sgr A$^*$.
This is supported by the observations summarized in Fig.~5.8 of
\citet{W_03_dust}.  }

\subsection{The crystalline fraction in the diffuse ISM}
\label{sec:crystalline}

We  determined the degree of crystallinity by adding the
mass absorption coefficients of some crystalline silicates to the sum
in Eq.~(\ref{eq:tau}). We have limited our search to the components
already identified to be present in astrophysical environments. These
are forsterite, clino- and ortho-enstatite, and diopside. For
consistency, we adopted the measurements obtained in one
{ experimental set-up}. Since diopside is only measured in one laboratory
\citep{KTS_00_diopside}, we thus used the mass absorption coefficient
($\kappa$) measurements of clino- and ortho-enstatite and forsterite
from the same laboratory
\citep{KTS_99_xsils}. The measurements were obtained from minerals crushed in 
a mortar, leading to non-spherical grain shapes. The grains larger
than 0.5 $\mu$m in size were removed, and the remaining grains were
incorporated in a KBr pellet. When using the mass absorption
coefficients $\kappa$ from these experiments we assume that the grain
shape and size distribution of the sample resembles that of the
crystalline silicates in the ISM.

\begin{figure}
\plotone{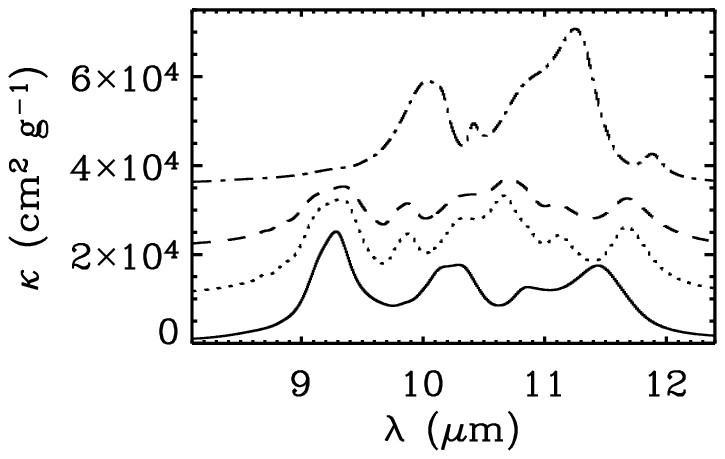}
\caption{{ Mass absorption coefficients $\kappa$ of the considered crystalline silicates.
The values for diopside \citep[solid line;][]{KTS_00_diopside},
clino-enstatite (dotted, offset by $1 \times 10^4$ cm$^{2}$ g$^{-1}$), ortho-enstatite (dashed, offset by $2 \times 10^4$ cm$^{2}$ g$^{-1}$) and forsterite
\citep[dash-dotted, offset by $3.5 \times 10^4$ cm$^{2}$ g$^{-1}$;][]{KTS_99_xsils} are indicated. These mass absorption coefficients are used to determine the 
the crystalline fraction in the studied interstellar line-of-sight
(see Fig.~\ref{fig:tau}).}  }
\label{fig:xsils}
\end{figure}

The small residues remaining when the fit with only amorphous
silicates was performed, suggest that there is not much room for
crystalline silicates left. The mass fraction of crystalline silicates
is probably considerably less than 3\% (the level of the residues
compared to the total optical depth) because the resonances produced
by crystalline silicates in the 10 $\mu$m region are not only much
narrower (see Fig.~\ref{fig:xsils}) but also intrinsically much
stronger than the resonances due to amorphous silicates.

The $\kappa$ values of all four species of crystalline silicates are
{ included in} Eq.~(\ref{eq:tau}). We have { minimized} $\chi^2$
and find that the best fit is given by a crystalline over amorphous
fraction of 0.2 \% by mass. The composition of crystalline silicates
found in this best fit is dominated mostly by forsterite, while $\sim$
1/6 of the crystalline silicates are in the form of either diopside or
enstatite. It is hard to distinguish between the contributions of the
various crystalline pyroxenes, as their strongest resonance at $\sim$
9.2 $\mu$m largely overlaps. On the other hand, the olivines are
easily distinguished from the pyroxenes, because of the presence of a
strong resonance at $\sim$ 11.2 $\mu$m.

In order to determine the amount of each individual crystalline dust
component that could be present, we have added them separately to the
amorphous mixture and determined the best $\chi^2$ fit. The $F$-test
was performed to determine the significance of adding these different
dust components. The results are shown in Table~\ref{tab:adding}. We
find that the most significant improvement is achieved when forsterite
is added to a completely amorphous mixture, resulting in an $F$ value
of 900. A smaller improvement, but still significant, is achieved by
adding diopside to a completely amorphous composition. Adding
ortho-enstatite does not lead to a very significant improvement, which
can be concluded from the small $F$ value (see
Sect.~\ref{sec:method}). Adding clino-enstatite does not improve the
fit at all. When diopside is added to a mixture of amorphous silicates
and forsterite also only a minor improvement is achieved.

To study the significance of the result that 0.2\% of the silicates
are crystalline, we have constructed a variety of dust compositions,
ranging from 0.1\% to 3\% in degree of crystallinity $x$. We used a
mass ratio of 5.6:1 for amorphous olivine:pyroxene, and a ratio of 6:1
for forsterite:(enstatite+diopside).  We assumed that diopside and
enstatite were present in equal amounts. For each of these
crystallinities, the best fits by minimizing the $\chi^2$ and
$\chi^2_\nu$ value are obtained, and compared with the fit for 0\%
crystallinity, by performing the $F$-test. The results are given in
Table~\ref{tab:chisq}. For $x \le 0.4$\%, the $\chi_\nu^2$ fit has
improved compared to the purely amorphous dust composition. Since all
the $F_\chi \gg 1$, the improvement of the fit is significant. For $x
> 0.4$\%, the fit deteriorates quickly with increasing $x$, and the
resulting $F$ values are negative. Examination of the residuals allows
us to firmly determine that the degree of crystallinity of silicates
in the diffuse ISM is $\le 0.4$\%. In Fig.~\ref{fig:tau} some sample
residuals are shown. For $x \ge 0.5$\% for example, the overabundance
of crystalline silicates clearly appears in the residuals and in the
model spectrum. Hence we derive that the degree of crystallinity in
the diffuse interstellar medium is 0.2\% $\pm$ 0.2\%.

Although $x = 0.2$\% gives the best fit to the optical depth in the 10
$\mu$m silicate feature, we would like to point out that this number
may be an upper limit as well. It is mostly determined by the 11.2
$\mu$m forsterite feature. Other possible interstellar dust components
such as PAHs, carbonates, { water ice and SiC} have strong
resonances near this wavelength too, which may account for part of the
optical depth at that particular wavelength, thus decreasing the
contribution and mass fraction of crystalline silicates.

From this analysis it becomes clear that forsterite may be present in
the diffuse interstellar medium, at a mass fraction of at most
0.2\%. The other crystalline components only contribute in minor
amounts.

\subsection{Other dust components}

{ Besides crystalline silicates, several other dust components have
resonances overlapping with the wavelength range covered by the
amorphous silicate absorption feature. Of high astrophysical interest
is silicon carbide (SiC), a dust component commonly produced by
carbon-rich AGB stars. In a similar fashion as described here, the
mass of SiC with respect to silicates can be determined, and compared to
the dust mass ratio produced by carbon-stars versus
oxygen-stars. Indeed,
\citet{WDM_90_GC} were able to determine that the abundance of Si
atoms in interstellar SiC was less than 5\% of the abundance of Si in
interstellar silicates, which translates into a mass fraction taken by
SiC of $< 1.7$\% of the total Si-containing containing dust.  This result
is based on low resolution
7.5 -- 13.5 $\mu$m spectroscopy of Galactic Center sources
\citep{RA_85_extinction}. \citet{WDM_90_GC} conclude that SiC probably is destroyed
upon ejection into the interstellar medium on a time scale of $\sim 5
\times 10^{7}$ years.

We have used the optical properties presented by \citet{LD_93_SiC} to
calculate how many spherical SiC grains with radius 0.1 $\mu$m could
be present in the line-of-sight towards the Galactic Center. A strong
resonance at $\sim$10.6 $\mu$m may be used to determine the abundance
of SiC in the diffuse ISM.  We find that adding the opacity of SiC to
a mixture of amorphous silicates or to a mixture of amorphous and
crystalline silicates only deteriorates the fit, leading to negative
$F_\chi$-values. Our conclusion is that 0.1 $\mu$m sized spherical SiC
grains make up $< 0.1$\% of the interstellar Si-bearing dust mass.  
This is a factor of 17 better than the previous determination based
on ground-based 10 $\mu$m spectra \citep{WDM_90_GC}. Analysis of this
result requires the use of reliable optical properties of SiC, which
are currently unavailable
\citep{SHB_99_SiC}.
In a future study we will re-address this issue (F.~Kemper \&
A.~Speck, \emph{in prep.}). }

\section{Silicate processing in the diffuse ISM}
\label{sec:processing}

\subsection{Injection of circumstellar silicates}

It is generally believed that AGB stars and red supergiants dominate
the production of oxygen-rich dust in the Galaxy \citep[see e.g.~][and
references herein]{W_03_dust}.  Most oxygen-rich dust produced by
these stars is in the form of silicates. The density of M supergiants
is found to be 1--2 kpc$^{-2}$ \citep{JK_90_supergiants} projected on
the galactic plane, while the density of AGB stars is $\sim$25
kpc$^{-2}$
\citep{JK_89_solarneighbourhood}. Almost all supergiants and about
half of the AGB stars produce silicate dust. Since OH/IR stars occur
10 times less frequent than Miras \citep{H_96_review}, we can infer
that the density of Miras is $\sim$11.4 kpc$^{-2}$, while the density
of OH/IR stars is $\sim$1.1 kpc$^{-2}$. Adopting a typical mass-loss
rate of 10$^{-7}$ $M_{\odot}$ yr$^{-1}$ for the Miras and 10$^{-4}$
$M_{\odot}$ yr$^{-1}$ for the M supergiants and OH/IR stars \citep[see
e.g.][and references herein]{H_96_review} and assuming that the
dust-to-gas ratio is 0.01 we are able to determine the dust injection
rate (Table~\ref{tab:replenish}). We assume that all oxygen-rich
dust produced by these stars is in the form of silicates.  The
crystalline fraction of AGB stars is relatively well known. It is
estimated that about 10 \% of the silicates around OH/IR stars are
crystalline, while for the Miras only an upper limit of 40\% of the
crystallinity can be derived \citep{KWD_01_xsilvsmdot}. The degree of
crystallinity in the spectra of M supergiants is less well known, in
part because most of those dust shells are optically thin and may be
able to hide a large amount of crystalline silicates in a similar way
as described in \citet{KWD_01_xsilvsmdot}. We use the supergiant AFGL
4106 as an example and adopt the derived crystallinity of 15--20\%
\citep{MWT_99_afgl4106} as a typical value. Hence, the crystallinity of the silicates
in the combined stellar ejecta will be 11--18\%, where 4--5\%
originates from OH/IR stars and 7--13\% from M supergiants. The
contribution from Miras is negligible (Table~\ref{tab:replenish}). One has to realize that most of these
estimates are based on a few objects, and the degree of crystallinity
of stellar ejecta is inherently an uncertain number.

The degree of crystallinity of silicates in the ejecta of stars
(11--18\%) { contributing to the} interstellar
dust { budget} is significantly different from the crystallinity observed in
interstellar silicates (0.2\% $\pm$ 0.2\%). We will explore two
possible explanations for this discrepancy. We mainly focus on the
possibility that an amorphization process occurs on time scales
significantly shorter than the destruction time scales, in order to
decrease the degree of crystallinity, { specifically} amorphization by
cosmic ray hits { or particle bombardment} (Sect.~\ref{sec:amorphization}).  In addition, we
consider the possibility that the crystalline fraction in stellar
ejecta is diluted by a yet neglected source of amorphous silicates,
greatly decreasing the average crystallinity of newly synthesized dust
(Sect.~\ref{sec:dilution}). We { will} discuss the dust production by
supernovae (SNe), recently suggested to be a significant source of
galactic dust \citep{DEI_03_SNdust}.

\subsection{Amorphization in the diffuse ISM}
\label{sec:amorphization}

In previous studies \citep[e.g.][]{JTH_94_graindestruction} the grain
destruction rates and dust life time in the interstellar medium have
been estimated. However, to date the rates and time scales involved
with crystallization and amorphization in interstellar conditions have
not yet been addressed. In this work, we will derive an expression for
the grain amorphization rate, and we will compare this to laboratory
experiments on the amorphization of grains in astrophysical
conditions.

The total mass of crystalline and amorphous silicates { in the ISM} can be written
in the form of coupled differential equations:

\begin{equation}
\label{eq:mass}
\left\{ \begin{array}{l} 
\frac{dM_{\mathrm{X}}}{dt} = x_{\ast}\dot{M}_{\ast} - k_1 M_{\mathrm{X}} + k_2 M_{\mathrm{A}} - k_3 M_{\mathrm{X}}\\
\frac{dM_{\mathrm{A}}}{dt} = (1 - x_{\ast})\dot{M}_{\ast} + k_1 M_{\mathrm{X}} - k_2 M_{\mathrm{A}} - k_3 M_{\mathrm{A}}
\end{array} \right.
\end{equation}

where $M_{\mathrm{X}}$ and $M_{\mathrm{A}}$ represent the crystalline
and amorphous silicate mass in the ISM respectively. $\dot{M}_{\ast}$
is the injection rate of stellar silicates assumed to be constant over
time, with $x_\ast$ the mass fraction of stellar silicates that are
crystalline, which is also constant. The amorphization rate is given
by $k_1$, the crystallization rate by $k_2$ and the destruction rate
by $k_3$. Under interstellar conditions $k_2$ will be very small, and
therefore we assume it to be 0.  We assume that at the current time
equilibrium is reached, which implies the stationary solution, i.e.

\begin{equation}
\label{eq:stationary}
\left\{ \begin{array}{l} 
\frac{dM_{\mathrm{X}}}{dt} = 0\\
\frac{dM_{\mathrm{A}}}{dt} = 0
\end{array} \right.
\end{equation}

By combining Eqs.~(\ref{eq:mass}) and (\ref{eq:stationary}) it is
possible to determine $k_1$ and $k_3$, assuming $k_2 = 0$. We find
that

\begin{equation}
k_1 = k_3 \frac{x_\ast - x_{\mathrm{ISM}}}{x_{\mathrm{ISM}}}
\end{equation}

and

\begin{equation}
k_3 = \frac{\dot{M_\ast}}{M_{\mathrm{X}}+M_{\mathrm{A}}}
\end{equation}

where $x_{\mathrm{ISM}}$ is the mass fraction of silicates in the ISM that
has a crystalline lattice structure, given by $x_{\mathrm{ISM}} =
M_{\mathrm{X}}/(M_{\mathrm{A}}+ M_{\mathrm{X}})$.

For silicate grains the destruction rate $k_3$ is $\sim 2 \times
10^{-9}$ yr$^{-1}$, corresponding to a dust life time of $4 \times
10^8$ yr
\citep{JTH_94_graindestruction,JTH_96_grainsize}, although 
\citet{T_98_lifecycle} derives a significantly shorter life time
of $6 \times 10^7$ yr, using observed depletions. Adopting the longest
time scale of $4 \times 10^8$ yr as a conservative choice, the
measured crystallinities $x_{\mathrm{ISM}} = 0.002$ and $x_\ast =
0.15$ require an amorphization rate $k_1$ of $1.5 \times 10^{-7}$
yr$^{-1}$ to reach a final crystallinity of 0.2\%. This corresponds to
an amorphization time scale of $\sim$ 5 Myr. Because the upper limit
to the crystallinity of the silicates in the ISM is 0.4\%, the
amorphization rate should be faster than $7.3 \times 10^{-8}$
yr$^{-1}$, corresponding to a time scale of $\sim$ 9 Myr or
shorter. Better determinations of the degree of crystallinity of the
silicates in stellar ejecta may yield a change in the amorphization
time scale. The time scales derived here should therefore be treated
only as estimates.

The most likely process to cause amorphization on time scales of
$\sim$ 5 Myr are ion bombardments of incident { particles} on
crystalline grains. Studies of dust grains collected by the Apollo
astronauts on the surface of the moon suggest that amorphization by
ion bombardments occurs. These studies have revealed that most grains
are surrounded by a 60--200 nm thick amorphous rims.  It is suggested
that these rims are caused by solar wind radiation damage, during the
relatively short period (5,000 -- 10,000 years) that each particular
grain was exposed to it before the grain was covered by other
extralunar grains deposited on the moon's surface
\citep{BCJ_80_lunar}. Although other mechanisms for the amorphization
have been suggested, such as the deposition of vapor phase materials
after the impact of a meteorite on the surface of the moon
\citep{KM_93_regolith}, it is thought that at least part of the
amorphous material found of the moon is the result of damage by
particles from the solar wind \citep{KM_97_rim}.

To date several laboratory studies have investigated the effect of ion
bombardments on crystalline grains
\citep[e.g.][]{D_77_irradiation,B_94_anomalousIDP,WWE_98_amorphization,DCL_01_He+,CDC_02_amorphization,JFS_03_bombardment,BSB_03_amorphisation},
measuring the required dose for amorphization as a function of energy
and element used. In general, the dose $D_n$ (eV cm$^{-3}$) can be
written as

\begin{equation}
\label{eq:dose}
D_n = \Phi S_n
\end{equation}

where $\Phi$ (cm$^{-2}$) is the particle fluence and $S_n$ (eV
cm$^{-1}$) the stopping power. From the compilation of experimental
data presented by \citet{BSB_03_amorphisation} it becomes clear that
for 30-60 keV ions 80\% disorder is achieved for doses over $20 \times
10^{23}$ eV cm$^{-3}$. { In this regime, the experiments are
dominated by nuclear stopping power.} The stopping powers $S_n$ for
the ions considered are: $S_n (30$ keV H$^+)\approx 0.08$ eV
\AA$^{-1}$; $S_n (30$ keV He$^+)\approx 0.8$ eV \AA$^{-1}$; $S_n (60$
keV Ar$^{2+})\approx 90$ eV \AA$^{-1}$
\citep{BSB_03_amorphisation}.  The Ar$^{2+}$ abundance is not known for these
low energy cosmic rays, but using intermediate and high energy
measurements, as well as cosmic abundances of the elements, we
estimate that the abundance with respect to He is
$\left[\textrm{Ar/He}\right] \approx -4$ \citep[see e.g.~][]{W_88_CR}.
Although Ar$^{2+}$ is not very abundant in cosmic rays it actually
deposits a large amount of energy over a short distance. The particle
flux necessary to achieve the required dose for amorphization follows
from Eq.~(\ref{eq:dose}), and we find that $\Phi = 2.2 \times 10^{14}$
cm$^{-2}$, to be collected over less than $\sim$9 Myr. Extrapolating
using Eq.~(3) from
\citet{BSB_03_amorphisation}, we { obtain an estimate that the intensity 
of particles with energies of}
60 keV equals 1 particle cm$^{-2}$ s$^{-1}$ sr$^{-1}$, { although 
this formula has not been verified for this energy range.}  
Integrating over an entire sphere
and over the maximum amorphization time scale of 9 Myr, we find that a
grain has seen $3.5 \times 10^{15}$ particles cm$^{-2}$, containing a
dose of $<3.5 \times 10^{11}$ Ar$^{2+}$ ions cm$^{-2}$. This is not
sufficient to amorphitize a crystalline grain by means of Ar$^{2+}$
bombardment. It has been suggested, that the more abundant
Fe$^{2+}$-ions ($\left[\textrm{Fe/He}\right] \approx -3$, see
\citet{W_88_CR}) might be able to explain the amorphization. 
The stopping power of a 60 keV Fe$^{2+}$ ion is 75 eV \AA$^{-1}$
(E.~Bringa,
\emph{priv.comm.}), however, the required dose of Fe$^{2+}$ for
amorphization remains yet to be experimentally determined, although
calculations of the amorphization process are well on their way
\citep{BJ_02_nanograins,BJ_04_ionerosion}. Processing by the
abundant C,N,O-ions may contribute to the amorphization as well.  {
Finally, as stated before, the ion flux in the keV regime is not
known, and remains subject to further study. The amorphization time
scale may in fact provide a constraint on the ion flux in the keV
regime.}

\subsection{Dilution of stellar ejecta}
\label{sec:dilution}

A significant source of silicate dust may be the production by
supernovae. Recently, it has been suggested that type II supernovae
\citep{DEI_03_SNdust} and type Ib supernovae
\citep{MDE_03_kepler} can produce on the order of a solar mass of dust
per supernova, which is much higher than what is generally accepted
\citep[][see also Table~\ref{tab:injection}]{W_03_dust}. 
These new results are based on SCUBA/JCMT photometry at 450 and 850
$\mu$m of Cas A and Kepler's supernova remnant respectively.  The
submm photometry requires a cold dust component (in addition to the
warm dust component traced by IRAS photometry) to explain the spectral
energy distribution. Although the authors argue that this dust is
produced by the supernovae, it may also be swept up interstellar dust.
Depending on the choice of FIR/submm dust emissivities it is estimated
that Cas A is surrounded by 2-4 $M_\odot$ of cold dust, while Kepler's
supernova remnant contains $\sim 1 \, M_\odot$ in the form of cold
dust. It is debatable whether two data points provide sufficient
information to constrain the complex physical environment in a
supernova remnant, where the number of free parameters is only
constrained by one's imagination, and describe properties such as
geometry, density distribution, clumpiness, grain properties, dust
composition and the interaction with the surrounding interstellar
medium. { Indeed, \citet{D_04_CasA} shows that a much smaller dust mass
in the form of metallic iron needles explains the same submm data points.}
The galactic dust production rate by SNe is estimated to
be $(7-15) \times 10^{-3} \, M_\odot$ yr$^{-1}$
\citep{DEI_03_SNdust,MDE_03_kepler}, which is dominated by the cold
dust. The authors have derived this rate by scaling the dust
production rates for type II supernovae \citep{TF_01_typeII} with the
dust condensation efficiency determined for Cas A, assuming it had a
30 $M_\odot$ progenitor, and applying a supernova rate, which arises
from standard values of the initial mass function and star formation
rate (L.~Dunne,~\emph{priv.comm.}). When this supernova dust
production rate is compared with the production rate from other stars
\citep[$\sim 5 \times 10^{-3} \, M_\odot$ yr$^{-1}$; ][]{W_03_dust},
one can conclude that 60--75 \% of the galactic dust originates from
supernovae.

Unfortunately, the two SCUBA data points used for this determination
do not yield any information on the composition of this cold dust
component. The presence of a 22 $\mu$m feature in the warm dust
component of Cas A is reported \citep{ADM_99_SN22um}, which is
attributed to a Mg-protosilicate. Mg-protosilicate is a MgO/SiO$_2$
gel formed in a solution of Na-silicate and MgCl$_2$
\citep{D_74_protosilicate}, which can be seen as a Mg-rich amorphous silicate. 

In order to estimate an upper limit on the dilution, we assume that
all dust produced by supernova comes out in the form of amorphous
silicates. Thus, if we adopt that 60--75 \% of the galactic dust
originates from supernova, and that the remaining 25--40 \% produced
by evolved stars has a degree of crystallinity of 11--18 \%, one
easily sees that in case of the most efficient dilution, the combined
stellar ejecta still have a degree of crystallinity of 3--7\%, which
is still more than an order of magnitude larger than the observed
crystallinity of (0.2 \% $\pm$ 0.2\%) observed for silicates in the
diffuse interstellar medium.  Clearly, dilution by supernova dust does
not fully explain the low degree of crystallinity observed in the
diffuse interstellar medium, although it may contribute to some
extent.

We will not consider dilution by silicates formed in the diffuse ISM.
In the low density environment of the diffuse interstellar medium, the
formation of amorphous silicates from gas phase molecules is unlikely,
although the high depletion of Si and high grain destruction rate
suggest that grain formation occurs
\citep{JTH_94_graindestruction,D_03_dust}. The observed difference in depletion
between cloud- and intercloud-regions in the ISM \citep[see
e.g.][]{SCS_94_abundance} show that dust destruction and formation
commonly occurs in the ISM \citep{T_98_lifecycle}.

\section{Comparison with presolar grains, young stars and the dense
interstellar medium}
\label{sec:comparison}

Crystalline silicates are reported to be present in the dense
interstellar environment, based on the identification of features in
the ISO spectra. While the presence of crystalline silicates in
circumstellar environments is well established, their presence in the
dense ISM remains subject to discussion.  Dense ISM crystalline
silicates were first reported to be present in Orion
\citep{CJL_00_xsil_in_orion}, but the features ascribed to
crystalline silicates in this work may very well be explained by a
combination of a PAH emission plateau, memory effects in the data and
incorrect band merging (F.~Kemper,~\emph{in
prep.}). \citet{OO_03_carbononions} report on the detection of
diopside in the Carina H{\sc ii} region, based on the identification
of the 60 $\mu$m feature by diopside. In other environments, other
identifications of this feature have been suggested, including water
ice around evolved stars
\citep{B_98_LWS_AGB}, dolomite in planetary nebulae
\citep{KJW_02_carbonates}, and recent measurements of melilite also
look promising as an identification of the 60 $\mu$m feature
\citep{CKT_03_melilite}.  Even though there are some indications that
crystalline silicates are present in dense clouds associated with young stellar objects, at most
1-2 \% of the silicates seen in the lines-of-sight towards protostars
can have a crystalline lattice structure
\citep{DJD_99_dustcomposition}. 

Crystalline silicates are often found the direct circumstellar environments
of pre-main-sequence stars and young main-sequence stars. They are
detected in debris disk object $\beta$ Pic
\citep{KFT_93_betapic}, in some T Tauri stars \citep{HKO_03_TTau} and
are quite common in the more massive Herbig Ae/Be stars
\citep[e.g.~][]{BMD_01_processing,MWB_01_haebe}. To date, no crystalline silicates have been
found in the younger, deeply embedded class 0 objects
\citep{DJD_99_dustcomposition}.

The lack of crystalline silicates in the diffuse interstellar medium
suggests that crystalline silicates found in environments associated
with star formation and dense clouds are formed or annealed
locally. There might have been an exchange of crystalline silicates
between dense clouds, cloud cores and circumstellar disks, but any
excursion into the diffuse ISM would make a crystalline silicate
amorphous on a very short time scale.

In the solar system, cometary dust particles are generally believed to
be the most pristine dust particles, which survived the formation of
the solar system without being processed. Infrared spectroscopy of
comets and laboratory analysis of cometary dust particles has revealed
that comets contain a considerable fraction of crystalline silicates,
usually around 30\%. The remainder of the silicates is found to be
amorphous \citep[][and references herein]{W_02_cometgrains}. Recent
model calculations, however, suggest that the amount of crystalline
silicates in comet Hale-Bopp may be significantly lower than
previously assumed (M.~Min,~\emph{in prep.}). The crystallization of
these cometary silicates must have occurred after the grain has left
the diffuse interstellar medium, either in the dense molecular cloud,
during the star formation process or in the solar nebula, and thus
cometary grains do not represent unaltered interstellar dust. Another
way to study cometary dust is by collecting interplanetary dust
particles (IDPs) of cometary origin in the stratosphere of the
Earth. Both the remote observations of comets and the study of IDPs
suggest that the crystalline fraction in comets is much larger than in
the interstellar medium. That suggests perhaps whole scale
recrystallization and mixing in the solar nebula { has occurred, in
which grains are evaporated and recondensed in the crystalline form}.

\citet{MKS_03_silicateIDP} have studied a
sample of nine anhydrous IDPs collected by NASA over a number of years
and were able to identify 1031 individual silicate subgrains. Six of
these grains have non-solar oxygen isotopic compositions, indicating a
presolar origin of these grains. The isotopic ratios { point to
formation} by red supergiants or AGB stars. It is important to realize
that grains with an isotopic composition that resembles the solar
composition may have condensed in the presolar nebula, but could also
originate from the prestellar nebulae of young stellar objects in the
vicinity of the solar nebula. The six grains of non-solar isotopic
composition therefore only represent a lower limit to the number of
presolar or interstellar grains in the sample. For three of the grains
with non-solar isotopic ratios the mineralogical composition has been
determined; two of them are found to be GEMS grains \citep[glasses
with embedded metals and sulfides, i.e.~amorphous
silicates;][]{B_94_anomalousIDP} and one turns out to be a forsterite
grain. From this one might conclude that 1/6 of the interstellar
grains are crystalline, which contradicts our result. The IDP study
clearly suffers from small number statistics, and their result will
become more significant if more data on the composition of presolar
grains can be acquired.

\section{Conclusions}
\label{sec:conclusions}

In this work, the results of a study on the crystallinity of silicates
in the diffuse ISM have been presented. It is known that crystalline
silicates are common in the circumstellar environment of both pre- and
post-main-sequence stars. It is generally accepted that the dust
particles found in the circumstellar environment of young stars
originate from evolved stars, and have arrived in their current
location after a long (several Gyr) residence time in the interstellar
medium. The puzzling lack of evidence of crystalline silicates in the
ISM, prompted us to set a firm upper limit on the crystallinity.

Studying the 10 $\mu$m silicate feature in absorption towards Sgr
A$^*$ we are able to determine that at most 0.4\% of the silicates in
the interstellar medium have a crystalline structure. The data are
best fitted with a degree of crystallinity of 0.2\%. In addition, we
have determined the composition of the amorphous silicate component in
the diffuse interstellar medium. We found that 84.9\% of the amorphous
grains are olivine (Mg$_{2x}$Fe$_{2(1-x)}$SiO$_4$) and 15.1\% are
pyroxene (Mg$_{x}$Fe$_{(1-x)}$SiO$_3$). The amorphous grains were
found to be spherical and smaller than $\sim 0.1$ $\mu$m in
radius. { Detailed analysis of the 10 $\mu$m feature indicates that
the pyroxenes are probably slightly Mg-rich (with $0.5 < x < 0.6$) and
the interstellar olivines may be Fe-rich (with $0.4 < x < 0.5$). }

By comparison of these results with the crystallinity of the silicates
produced by mass-losing stars, and the interstellar grain destruction
rate, we have determined the interstellar amorphization rate, and
found that crystalline silicates are effectively amorphitized in 5 Myr
to achieve a final crystallinity of 0.2\%, while an amorphization time
scale of 9 Myr is consistent with the determined upper limit of 0.4\%.
These numbers are only estimates, as the crystallinity of stellar
ejecta is not determined very accurately yet.

Amorphization by low energy { ion bombardment} has been explored as
an explanation for the low degree of crystallinity of silicates in the
ISM, but the results { are not yet} consistent with the observed
low degree of crystallinity.  In addition, dilution by other sources
of amorphous silicates, such as supernovae, are not sufficient to
explain the observed lack of crystalline silicates in the ISM.

We conclude that effectively all silicates in the diffuse interstellar
medium become amorphous on a very short time scale, compared to the
total residence time in the diffuse ISM. Hence, the crystalline
silicates found in the circumstellar environments of young stars, in
the solar system and in star formation regions have not survived the
ISM unaltered, but are probably crystallized locally. Both annealing
as well as evaporation and subsequent condensation seem to be
significant crystallization processes.

\begin{acknowledgements}
We thank Don Brownlee, Lindsay Keller and Scott Messenger for
organizing the workshop on cometary dust in astrophysics, held in
Crystal Mountain, WA, in August 2003. This work greatly benefited from
the fruitful interdisciplinary interactions at this meeting. 
{ We thank Angelle Tanner for making available the 12.4 $\mu$m map
of the Galactic Center region.}
FK wishes to thank Mark Logan for the inspiring discussions. 
Support for this work was provided by NASA through the Spitzer Fellowship
Program, under award 011 808-001.  The data presented was analysed
with the support of the Dutch ISO Data Analysis Centre (DIDAC) at the
Space Research Organisation Netherlands (SRON) in Groningen, the
Netherlands.
\end{acknowledgements}

\clearpage

\begin{deluxetable}{l c c c c c r}
\tablecaption{Details of the ISO SWS observations \label{tab:obs}}
\tablehead{\colhead{object} & \colhead{$\alpha$ (J2000)} & \colhead{$\delta$ (J2000)} & \colhead{AOT} & \colhead{speed} & \colhead{$t_{\mathrm{int}}$ (s)} & \colhead{obs.~date}}
\startdata
Sgr A$^{*}$ & 17 45 40.0 & $-$29 00 29 & 1 & 4 & 6528 & 19-feb-96 \\
GCS 3       & 17 46 14.8 & $-$28 49 34 & 1 & 3 & 3454 & 29-aug-96 \\
GCS 4       & 17 46 15.7 & $-$28 49 47 & 1 & 3 & 3454 & 8-sep-96\\
\enddata
\end{deluxetable}

\begin{deluxetable}{l l r r r c c c c c c}
\tablecaption{Improvements of the spectral fit achieved by adding new dust components. The improvement is considered significant when $F_\chi$ is large (see text). The first column gives the initial dust components, while the newly 
added dust component is given in the second column. The change in
$\chi^2$ is given in the third column, and the reduced $\chi^2_\nu$
corresponding to the best fit with the new dust component added is
presented in the fourth column. Column 5 gives the value for
$F_\chi$. Columns 6--11 give the relative abundance of each dust component
corresponding to the new best fit. Only adding forsterite or diopside
significantly improves the fit. The first line of data gives the best
values for a completely amorphous dust
composition. \label{tab:adding}}
\tablehead{\colhead{initial} & \colhead{add} & \colhead{$\Delta \chi^2$}&\colhead{$\chi_\nu^2$} & \colhead{$F_\chi$} & \colhead{oliv.}&\colhead{pyr.} & \colhead{forst.} & \colhead{diop.}& \colhead{c-enst.} & \colhead{o-enst.}}
\startdata
amorph. & --         & --    & 69.9 & --  & 84.9\% & 15.1\% & -- & -- & -- & --\\
\hline
amorph. & forsterite & 40007 & 44.5 & 900 & 82.7\% & 17.1\% & 0.2\% & -- & -- & --\\
amorph. & diopside   & 11024 & 62.9 & 175 & 84.3\% & 15.4\% & -- & 0.3\% & -- & --\\
amorph. & clino-enst. & $<$0 & -- & $<$0 & --  & --     & -- & -- & -- & --\\
amorph. & ortho-enst. & 552 & 69.6 & 7.9 & 84.7\% & 15.2\% & -- & -- & -- &0.07\% \\
\hline
am. + forst. & diopside & 139 & 44.4 & 3.1 & 82.7\% & 17.1\% & 0.2\% & 0.03\% & -- & -- \\
\enddata
\end{deluxetable}

\begin{deluxetable}{l c c}
\tablecaption{Overview of the fit results for various degrees of crystallinity. The first column represents the mass fraction $x$ of the silicates that is 
crystalline, the second column gives the minimized value of
$\chi_\nu^2$ determining the best fit. In case the $\chi_\nu^2$ has
decreased compared to the value corresponding to $x = 0$, the goodness
of the fit was evaluated using the $F$-test (column 3). For $F_\chi
\gg 1$ adding the crystalline silicates as an extra parameter really
improved the fit.  \label{tab:chisq}}
\tablehead{\colhead{$x$} & \colhead{$\chi^2_\nu$} & \colhead{$F_\chi$}}
\startdata
0.0 & 70 & -- \\
0.1 & 56 & 396\\ 
0.2 & 51 & 549\\
0.3 & 58 & 338\\
0.4 & 72 & $<$0 \\ 
0.5 & 96 & $<$0\\
0.7 & 168 & $<$0\\
1.0 & 331 &  $<$0\\
1.5 & 723 &  $<$0\\
2.0 & 1231 & $<$0\\
3.0 & 2470 &  $<$0\\
\enddata
\label{tab:f}
\end{deluxetable}

\begin{deluxetable}{l c c c c c}
\tablecaption{Contribution of dust producing stars to the crystalline silicate replenishment of the ISM. The first column gives the surface number density of each type of silicate dust producing star in the solar vicinity. The second column gives the dust mass loss rate, assuming a dust/gas ratio of 1/100. In the third column the mass fraction of crystalline silicates in the stellar outflow is given. Column 4 gives the mass injection rate in the ISM, which is in fact the product of the numbers in column 1 and 2. The fifth column then, gives the crystalline silicate injection into the ISM, relative to the total silicate injection rate. The total degree of crystallinity of the silicates injected into the ISM is the sum of the numbers in column 5. References: $^{\mathrm{a}}$\citet{JK_89_solarneighbourhood} $^{\mathrm{b}}$\citet{H_96_review} $^{\mathrm{c}}$\citet{JK_90_supergiants} $^{\mathrm{d}}$\citet{KWD_01_xsilvsmdot} $^{\mathrm{e}}$\cite{MWT_99_afgl4106} $^{\mathrm{f}}$\cite{W_03_dust} \label{tab:replenish}}

\tablehead{\colhead{type of star} & \colhead{$N$ (kpc$^{-2}$)} & \colhead{$\dot{M}_{\mathrm{dust}}$ ($M_{\odot}$ yr$^{-1}$)} & \colhead{$x$} & \colhead{$\dot{M}_{\mathrm{inj}}$ ($M_\odot$ yr$^{-1}$ kpc$^{-2}$)} & \colhead{($x$ $\dot{M}_{\mathrm{inj}}$)/$\dot{M}_{\mathrm{ISM}}$}}
\startdata
Miras         & 11.4$^{\mathrm{a,b}}$ & 10$^{-9}$ & $<$40\%$^{\mathrm{d}}$  & $1.14 \times 10^{-8}$ & $<$0.2\%\\ 
OH/IR stars   & 1.1$^{\mathrm{a,b}}$  & 10$^{-6}$ & 10\%$^{\mathrm{d}}$     & $1.1 \times 10^{-6}$  & 4--5\%\\ 
M Supergiants & 1--2$^{\mathrm{c}}$   & 10$^{-6}$ & 15--20\%$^{\mathrm{e}}$ & $1-2 \times 10^{-6}$  & 7--13\%\\ 
Supernovae    & n/a                   & n/a       & ?                       & $\sim 4 \times 10^{-7}$ $^
{\mathrm{f}}$& ?  \\
\enddata
\label{tab:injection}
\end{deluxetable}


\begin{thebibliography}{82}
\expandafter\ifx\csname natexlab\endcsname\relax\def\natexlab#1{#1}\fi

\bibitem[{Arendt {et~al.}(1999)Arendt, Dwek, \& Moseley}]{ADM_99_SN22um}
Arendt, R.~G., Dwek, E., \& Moseley, S.~H. 1999, \apj, 521, 234

\bibitem[{Baganoff {et~al.}(2003)Baganoff, Maeda, Morris, Bautz, Brandt, Cui,
  Doty, Feigelson, Garmire, Pravdo, Ricker, \& Townsley}]{BMM_03_xrayGC}
Baganoff, F.~K., Maeda, Y., Morris, M., {et~al.} 2003, \apj, 591, 891

\bibitem[{Barlow(1998)}]{B_98_LWS_AGB}
Barlow, M.~J. 1998, \apss, 255, 315

\bibitem[{Becklin {et~al.}(1978)Becklin, Matthews, Neugebauer, \&
  Willner}]{BMN_78_GCI}
Becklin, E.~E., Matthews, K., Neugebauer, G., \& Willner, S.~P. 1978, \apj,
  219, 121

\bibitem[{Bevington \& Robinson(1992)}]{BR_92_ftest}
Bevington, P.~R. \& Robinson, D.~K. 1992, Data reduction and error analysis for
  the physical sciences (second edition) (New York: McGraw-Hill)

\bibitem[{Bohren \& Huffman(1983)}]{BH_83_scattering}
Bohren, C.~F. \& Huffman, D.~R. 1983, Absorption and scattering of light by
  small particles (New York: Wiley)

\bibitem[{Borg {et~al.}(1980)Borg, Chaumont, Jouret, Langevin, \&
  Maurette}]{BCJ_80_lunar}
Borg, J., Chaumont, J., Jouret, C., Langevin, Y., \& Maurette, M. 1980, in
  Proc.~Conf.~Ancient Sun, ed. R.~O. Pepin, J.~A. Eddy, \& R.~B. Merrill,
  431--461

\bibitem[{Bouwman {et~al.}(2001)Bouwman, Meeus, {de Koter}, Hony, Dominik, \&
  Waters}]{BMD_01_processing}
Bouwman, J., Meeus, G., {de Koter}, A., {et~al.} 2001, \aap, 375, 950

\bibitem[{Bowey \& Adamson(2002)}]{BA_02_mineralogy}
Bowey, J.~E. \& Adamson, A.~J. 2002, \mnras, 334, 94

\bibitem[{Bradley(1994)}]{B_94_anomalousIDP}
Bradley, J.~P. 1994, \sci, 265, 925

\bibitem[{Bringa \& Johnson(2002)}]{BJ_02_nanograins}
Bringa, E.~M. \& Johnson, R.~E. 2002, Nuclear Instruments and Methods in
  Physics Research B, 193, 365

\bibitem[{Bringa \& Johnson(2004)}]{BJ_04_ionerosion}
---. 2004, \apj, 603, 159

\bibitem[{Brucato {et~al.}(2003)Brucato, Strazzulla, Baratta, \&
  Colangeli}]{BSB_03_amorphisation}
Brucato, J.~R., Strazzulla, G., Baratta, G., \& Colangeli, L. 2003, \aap, 413,
  395

\bibitem[{Carrez {et~al.}(2002)Carrez, Demyk, Cordier, Gengembre, Grimblot,
  {d'Hendecourt}, Jones, \& Leroux}]{CDC_02_amorphization}
Carrez, P., Demyk, K., Cordier, P., {et~al.} 2002, \mps, 37, 1599

\bibitem[{Cesarsky {et~al.}(2000)Cesarsky, Jones, Lequeux, \&
  Verstraete}]{CJL_00_xsil_in_orion}
Cesarsky, D., Jones, A.~P., Lequeux, J., \& Verstraete, L. 2000, \aap, 358, 708

\bibitem[{Chiar {et~al.}(1998)Chiar, Pendleton, Geballe, \&
  Tielens}]{CPG_98_hydrocarbon}
Chiar, J.~E., Pendleton, Y.~J., Geballe, T.~R., \& Tielens, A. G. G.~M. 1998,
  \apj, 507, 281

\bibitem[{Chiar {et~al.}(2001)Chiar, Tielens, Whittet, Schutte, Boogert, Lutz,
  {van Dishoeck}, \& Bernstein}]{CTW_01_GC}
Chiar, J.~E., Tielens, A. G. G.~M., Whittet, D. C.~B., {et~al.} 2001, \apj,
  537, 749

\bibitem[{Chihara {et~al.}(2003)Chihara, Koike, \&
  Tsuchiyama}]{CKT_03_melilite}
Chihara, H., Koike, C., \& Tsuchiyama, A. 2003, in Astrophysics of Dust, Estes
  Park, Colorado, May 26 - 30, 2003. Edited by Adolf N. Witt.

\bibitem[{Day(1974)}]{D_74_protosilicate}
Day, K.~L. 1974, \apjl, 192, L15

\bibitem[{Day(1977)}]{D_77_irradiation}
---. 1977, \mnras, 178, 49P

\bibitem[{{de Graauw} {et~al.}(1996){de Graauw}, Haser, Beintema, Roelfsema,
  {van Agthoven}, Barl, Bauer, Bekenkamp, Boonstra, Boxhoorn, Cot\'e, {de
  Groene}, {van Dijkhuizen}, Drapatz, Evers, Feuchtgruber, Frericks, Genzel,
  Haerendel, Heras, {van der Hucht}, {van der Hulst}, Huygen, Jacobs, Jakob,
  Kamperman, Katterloher, Kester, Kunze, Kussendrager, Lahuis, Lamers, Leech,
  {van der Lei}, {van der Linden}, Luinge, Lutz, Melzner, Morris, {van Nguyen},
  Ploeger, Price, Salama, Schaeidt, Sijm, Smoorenburg, Spakman, Spoon,
  Steinmayer, Stoecker, Valentijn, Vandenbussche, Visser, Waelkens, Waters,
  Wensink, Wesselius, Wiezorrek, Wieprecht, Wijnbergen, Wildeman, \&
  Young}]{GHB_96_SWS}
{de Graauw}, T., Haser, L.~N., Beintema, D.~A., {et~al.} 1996, \aap, 315, L49

\bibitem[{Demyk {et~al.}(2001)Demyk, Carrez, Leroux, Cordier, Jones, Borg,
  Quirico, Raynal, \& {d'Hendecourt}}]{DCL_01_He+}
Demyk, K., Carrez, P., Leroux, H., {et~al.} 2001, \aap, 368, L38

\bibitem[{Demyk {et~al.}(2000)Demyk, Dartois, Wiesemeyer, Jones,
  {d'Hendecourt}, {Jourdain de Muizon}, \& Heras}]{DDW_00_nearinfraredproblem}
Demyk, K., Dartois, E., Wiesemeyer, H., {et~al.} 2000, in ISO beyond the peaks:
  The 2nd ISO workshop on analytical spectroscopy, ESA-SP 456, 183--187

\bibitem[{Demyk {et~al.}(1999)Demyk, Jones, Dartois, Cox, \&
  {d'Hendecourt}}]{DJD_99_dustcomposition}
Demyk, K., Jones, A.~P., Dartois, E., Cox, P., \& {d'Hendecourt}, L. 1999,
  \aap, 349, 267

\bibitem[{Dorschner {et~al.}(1995)Dorschner, Begemann, Henning, J\"ager, \&
  Mutschke}]{DBH_95_glasses}
Dorschner, J., Begemann, B., Henning, T., J\"ager, C., \& Mutschke, H. 1995,
  \aap, 300, 503

\bibitem[{Draine(2003)}]{D_03_dust}
Draine, B.~T. 2003, \araa, 41, 241

\bibitem[{Draine \& Tan(2003)}]{DT_03_Xray}
Draine, B.~T. \& Tan, J.~C. 2003, \apj, 594, 347

\bibitem[{Dunne {et~al.}(2003)Dunne, Eales, Ivison, Morgan, \&
  Edmunds}]{DEI_03_SNdust}
Dunne, L., Eales, S., Ivison, R., Morgan, H., \& Edmunds, M. 2003, \nat, 424,
  285

\bibitem[{Dwek(2004)}]{D_04_CasA}
Dwek, E. 2004, \apj, in press

\bibitem[{Figer {et~al.}(1999)Figer, McLean, \& Morris}]{FMM_99_quintuplet}
Figer, D.~F., McLean, I.~S., \& Morris, M. 1999, \apj, 514, 202

\bibitem[{Fouks \& Schubert(1995)}]{FS_95_memory}
Fouks, B.~I. \& Schubert, J. 1995, \procspie, 2475, 487

\bibitem[{Habing(1996)}]{H_96_review}
Habing, H. 1996, \aapr, 7, 97

\bibitem[{Honda {et~al.}(2003)Honda, Kataza, Okamoto, Miyata, Yamashita, Sako,
  Takubo, \& Onaka}]{HKO_03_TTau}
Honda, M., Kataza, H., Okamoto, Y.~K., {et~al.} 2003, \apj, 585, L59

\bibitem[{{Iat{\`{\i}}} {et~al.}(2001){Iat{\`{\i}}}, {Cecchi-Pestellini},
  Williams, Borghese, Denti, Saija, \& Aiello}]{ICW_01_porous}
{Iat{\`{\i}}}, M.~A., {Cecchi-Pestellini}, C., Williams, D.~A., {et~al.} 2001,
  \mnras, 322, 749

\bibitem[{J\"ager {et~al.}(2003{\natexlab{a}})J\"ager, Dorschner, Mutschke,
  Posch, \& Henning}]{JDM_03_solgel}
J\"ager, C., Dorschner, J., Mutschke, H., Posch, T., \& Henning, T.
  2003{\natexlab{a}}, \aap, 408, 193

\bibitem[{J\"ager {et~al.}(2003{\natexlab{b}})J\"ager, Fabian, Schrempel,
  Dorschner, Henning, \& Wesch}]{JFS_03_bombardment}
J\"ager, C., Fabian, D., Schrempel, F., {et~al.} 2003{\natexlab{b}}, \aap, 401,
  57

\bibitem[{Jones {et~al.}(1996)Jones, Tielens, \& Hollenbach}]{JTH_96_grainsize}
Jones, A.~P., Tielens, A. G. G.~M., \& Hollenbach, D.~J. 1996, \apj, 469, 740

\bibitem[{Jones {et~al.}(1994)Jones, Tielens, Hollenbach, \&
  McKee}]{JTH_94_graindestruction}
Jones, A.~P., Tielens, A. G. G.~M., Hollenbach, D.~J., \& McKee, C.~F. 1994,
  \apj, 433, 797

\bibitem[{Jura \& Kleinmann(1989)}]{JK_89_solarneighbourhood}
Jura, M. \& Kleinmann, S.~G. 1989, \apj, 341, 359

\bibitem[{Jura \& Kleinmann(1990)}]{JK_90_supergiants}
---. 1990, \apjs, 73, 769

\bibitem[{Keller \& McKay(1993)}]{KM_93_regolith}
Keller, L.~P. \& McKay, D.~S. 1993, \sci, 261, 1305

\bibitem[{Keller \& McKay(1997)}]{KM_97_rim}
---. 1997, \gca, 61, 2331

\bibitem[{Kemper {et~al.}(2002)Kemper, J\"ager, Waters, Henning, Molster,
  Barlow, Lim, \& {de Koter}}]{KJW_02_carbonates}
Kemper, F., J\"ager, C., Waters, L. B. F.~M., {et~al.} 2002, \nat, 415, 295

\bibitem[{Kemper {et~al.}(2001)Kemper, Waters, {de Koter}, \&
  Tielens}]{KWD_01_xsilvsmdot}
Kemper, F., Waters, L. B. F.~M., {de Koter}, A., \& Tielens, A. G. G.~M. 2001,
  \aap, 369, 132

\bibitem[{Kessler {et~al.}(1996)Kessler, Steinz, Anderegg, Clavel, Drechsel,
  Estaria, Faelker, Riedinger, Robson, Taylor, \& {Xim\'enez de
  Ferr\'an}}]{KSA_96_ISO}
Kessler, M.~F., Steinz, J.~A., Anderegg, M.~E., {et~al.} 1996, \aap, 315, L27

\bibitem[{Kester(2003)}]{K_03_memory}
Kester, D. J.~M. 2003, in ESA SP-481: The Calibration Legacy of the ISO
  Mission, 243

\bibitem[{Knacke {et~al.}(1993)Knacke, Fajardo-Acosta, Telesco, Hackwell,
  Lynch, \& Russell}]{KFT_93_betapic}
Knacke, R.~F., Fajardo-Acosta, S.~B., Telesco, C.~M., {et~al.} 1993, \apj, 418,
  440

\bibitem[{Koike {et~al.}(2000)Koike, Tsuchiyama, Shibai, Suto, Tanab\'e,
  Chihara, Sogawa, Mouri, \& Okada}]{KTS_00_diopside}
Koike, C., Tsuchiyama, A., Shibai, H., {et~al.} 2000, \aap, 363, 1115

\bibitem[{Koike {et~al.}(1999)Koike, Tsuchiyama, \& Suto}]{KTS_99_xsils}
Koike, C., Tsuchiyama, A., \& Suto, H. 1999, in Proc.~32nd ISAS Lunar
  Planet.~Symp., 175--178

\bibitem[{Laor \& Draine(1993)}]{LD_93_SiC}
Laor, A. \& Draine, B.~T. 1993, \apj, 402, 441

\bibitem[{Leech {et~al.}(2003)Leech, Kester, Shipman, Beintema, Feuchtgruber,
  Heras, Huygen, Lahuis, Lutz, Morris, Roelfsema, Salama, Schaeidt, Valentijn,
  Vandenbussche, Wieprecht, \& {de Graauw}}]{LKS_02_ISOSWS}
Leech, K., Kester, D., Shipman, R., {et~al.}, eds. 2003, {The ISO Handbook,
  Volume V: SWS - The Short Wavelength Spectrometer}

\bibitem[{Li \& Draine(2001)}]{LD_01_silicate}
Li, A. \& Draine, B.~T. 2001, \apjl, 550, L213

\bibitem[{Lutz {et~al.}(1996)Lutz, Feuchtgruber, Genzel, Kunze, Rigopoulou,
  Spoon, Wright, Egami, Katterloher, Sturm, Wieprecht, Sternberg, Moorwood, \&
  {de Graauw}}]{LFG_96_gc}
Lutz, D., Feuchtgruber, H., Genzel, R., {et~al.} 1996, \aap, 315, L269

\bibitem[{Mathis(1998)}]{M_98_porous}
Mathis, J.~S. 1998, \apj, 497, 824

\bibitem[{Mathis {et~al.}(1977)Mathis, Rumpl, \& Nordsieck}]{MRN_77_grainsize}
Mathis, J.~S., Rumpl, W., \& Nordsieck, K.~H. 1977, \apj, 217, 425

\bibitem[{Meeus {et~al.}(2001)Meeus, Waters, Bouwman, {van den Ancker},
  Waelkens, \& Malfait}]{MWB_01_haebe}
Meeus, G., Waters, L. B. F.~M., Bouwman, J., {et~al.} 2001, \aap, 365, 476

\bibitem[{Messenger {et~al.}(2003)Messenger, Keller, Stadermann, Walker, \&
  Zinner}]{MKS_03_silicateIDP}
Messenger, S., Keller, L.~P., Stadermann, F.~J., Walker, R.~M., \& Zinner, E.
  2003, \sci, 300, 105

\bibitem[{Molster {et~al.}(2002)Molster, Waters, Tielens, \&
  Barlow}]{MWT_02_xsilI}
Molster, F.~J., Waters, L. B. F.~M., Tielens, A. G. G.~M., \& Barlow, M.~J.
  2002, \aap, 382, 184

\bibitem[{Molster {et~al.}(1999)Molster, Waters, Trams, {Van Winckel}, Decin,
  {van Loon}, J\"ager, Henning, K\"aufl, {de Koter}, \&
  Bouwman}]{MWT_99_afgl4106}
Molster, F.~J., Waters, L. B. F.~M., Trams, N.~R., {et~al.} 1999, \aap, 350,
  163

\bibitem[{Moneti {et~al.}(2001)Moneti, Stolovy, Blommaert, Figer, \&
  Najaro}]{MSB_01_quintuplet}
Moneti, A., Stolovy, S., Blommaert, J. A. D.~L., Figer, D.~F., \& Najaro, F.
  2001, \aap, 366, 106

\bibitem[{Morgan {et~al.}(2003)Morgan, Dunne, Eales, Ivison, \&
  Edmunds}]{MDE_03_kepler}
Morgan, H.~L., Dunne, L., Eales, S.~A., Ivison, R.~J., \& Edmunds, M.~G. 2003,
  \apjl, 597, L33

\bibitem[{Onaka \& Okada(2003)}]{OO_03_carbononions}
Onaka, T. \& Okada, Y. 2003, \apj, 585, 872

\bibitem[{Pendleton {et~al.}(1994)Pendleton, Sandford, Allamandola, Tielens, \&
  Sellgren}]{PSA_94_GC}
Pendleton, Y.~J., Sandford, S.~A., Allamandola, L.~J., Tielens, A. G. G.~M., \&
  Sellgren, K. 1994, \apj, 437, 683

\bibitem[{Press {et~al.}(1992)Press, Teukolsky, Vetterling, \&
  Flannery}]{PTV_92_recipes}
Press, W.~H., Teukolsky, S.~A., Vetterling, W.~T., \& Flannery, B.~P. 1992,
  Numerical recipes in Fortran 77 (second edition) (Cambridge University Press)

\bibitem[{Rieke {et~al.}(1989)Rieke, Rieke, \& Paul}]{RRP_89_gc}
Rieke, G.~H., Rieke, M.~J., \& Paul, A.~E. 1989, \apj, 336, 752

\bibitem[{Roche \& Aitken(1984)}]{RA_84_WCWR}
Roche, P.~F. \& Aitken, D.~K. 1984, \mnras, 208, 481

\bibitem[{Roche \& Aitken(1985)}]{RA_85_extinction}
---. 1985, \mnras, 215, 425

\bibitem[{Sofia {et~al.}(1994)Sofia, Cardelli, \& Savage}]{SCS_94_abundance}
Sofia, U.~J., Cardelli, J.~A., \& Savage, B.~D. 1994, \apj, 430, 650

\bibitem[{Speck {et~al.}(1999)Speck, Hofmeister, \& Barlow}]{SHB_99_SiC}
Speck, A.~K., Hofmeister, A.~M., \& Barlow, M.~J. 1999, \apjl, 513, L87

\bibitem[{Tanner {et~al.}(2002)Tanner, Ghez, Morris, Becklin, Cotera, Ressler,
  Werner, \& Wizinowich}]{TGM_02_GC}
Tanner, A., Ghez, A.~M., Morris, M., {et~al.} 2002, \apj, 575, 860

\bibitem[{Tielens(1998)}]{T_98_lifecycle}
Tielens, A. G. G.~M. 1998, \apj, 499, 267

\bibitem[{Tielens {et~al.}(1996)Tielens, Wooden, Allamandola, Bregman, \&
  Witteborn}]{TWA_96_GC}
Tielens, A. G. G.~M., Wooden, D.~H., Allamandola, L.~J., Bregman, J., \&
  Witteborn, F.~C. 1996, \apj, 461, 210

\bibitem[{Todini \& Ferrara(2001)}]{TF_01_typeII}
Todini, P. \& Ferrara, A. 2001, \mnras, 325, 726

\bibitem[{Vriend(1999)}]{V_99_10micronGC}
Vriend, W.~J. 1999, Master's thesis, Kapteijn Institute, Groningen University,
  The Netherlands

\bibitem[{Waelkens {et~al.}(1996)Waelkens, Waters, {de Graauw}, Huygen,
  Malfait, Plets, Vandenbussche, Beintema, Boxhoorn, Habing, Heras, Kester,
  Lahuis, Morris, Roelfsema, Salama, Siebenmorgen, Trams, {van der Bliek},
  Valentijn, \& Wesselius}]{WWD_96_xsilsyoung}
Waelkens, C., Waters, L. B. F.~M., {de Graauw}, M.~S., {et~al.} 1996, \aap,
  315, L245

\bibitem[{Wang {et~al.}(1998)Wang, Wang, Ewing, \&
  Doremus}]{WWE_98_amorphization}
Wang, S.~X., Wang, L.~M., Ewing, R.~C., \& Doremus, R.~H. 1998, \jncs, 238, 198

\bibitem[{Waters {et~al.}(1996)Waters, Molster, {de Jong}, Beintema, Waelkens,
  Boogert, Boxhoorn, de~Graauw, Drapatz, Feuchtgruber, Genzel, Helmich, Heras,
  Huygen, Izumiura, Justtanont, Kester, Kunze, Lahuis, Lamers, Leech, Loup,
  Lutz, Morris, Price, Roelfsema, Salama, Schaeidt, Tielens, Trams, Valentijn,
  Vandenbussche, {van den Ancker}, {van Dishoeck}, {van Winckel}, Wesselius, \&
  Young}]{WMJ_96_mineralogy}
Waters, L. B. F.~M., Molster, F.~J., {de Jong}, T., {et~al.} 1996, \aap, 315,
  L361

\bibitem[{Wefel(1988)}]{W_88_CR}
Wefel, J.~P. 1988, in Genesis and propagation of cosmic rays, ed. M.~M. Shapiro
  \& J.~P. Wefel (D. Reidel Publishing Company), 1--40

\bibitem[{Whittet(2003)}]{W_03_dust}
Whittet, D. C.~B. 2003, Dust in the galactic environment (second edition)
  (London: Institute of Physics Publishing)

\bibitem[{Whittet {et~al.}(1990)Whittet, Duley, \& Martin}]{WDM_90_GC}
Whittet, D. C.~B., Duley, W.~W., \& Martin, P.~G. 1990, \mnras, 244, 427

\bibitem[{Witt {et~al.}(2001)Witt, Smith, \& Dwek}]{WSD_01_Xray}
Witt, A.~N., Smith, R.~K., \& Dwek, E. 2001, \apjl, 550, L201

\bibitem[{Wooden(2002)}]{W_02_cometgrains}
Wooden, D.~H. 2002, \emp, 89, 247

\end{thebibliography}
\end{document}